\documentclass[twocolumn]{aastex63}

\newcommand{\Hnaut}{${\rm H}_{\rm 0}$}
\newcommand{\SDSSfiveim}{SDSS\,J1004+4112}
\newcommand{\SDSSthreeim}{SDSS\,J1029+2623}
\newcommand{\SDSSsixim}{SDSS\,J2222+2745}
\newcommand{\HST}{{HST}}
\newcommand{\fullHST}{\textit{Hubble Space Telescope}}
\newcommand{\unitsHnaut}{km s$^{-1}$ Mpc$^{-1}$}

\newcommand{\numConSDSSfiveim}{{78}}
\newcommand{\numParamsSDSSfiveim}{{27}}
\newcommand{\numConSDSSthreeim}{{48}}
\newcommand{\numParamsSDSSthreeim}{{33}}
\newcommand{\numConSDSSsixim}{{32}}
\newcommand{\numParamsSDSSsixim}{{31}}
\newcommand{\Lenstool}{{\tt{Lenstool}}}
\newcommand{\z}{\textit{z}}

\newcommand{\HfiveimABbest}{17.4}
\newcommand{\HfiveimAB}{56.1 $\pm$ 60.2}
\newcommand{\HfiveimCAbest}{66.4}
\newcommand{\HfiveimCA}{56.3 $\pm$ 19.1}
\newcommand{\HfiveimADbest}{55.5}
\newcommand{\HfiveimAD}{70.4 $\pm$ 11.4}

\newcommand{\HthreeimBAbest}{99.7}
\newcommand{\HthreeimBA}{97.0 $\pm$ 42.4}

\newcommand{\HsiximABbest}{89.8}
\newcommand{\HsiximAB}{104.9 $\pm$ 19.1}
\newcommand{\HsiximCAbest}{69.3}
\newcommand{\HsiximCA}{70.3 $\pm$ 17.4}

\newcommand{\Htot}{74.1 $\pm$ 8.0}
\newcommand{\Htotvalue}{74.1}
\newcommand{\Htoterr}{8.0}

\usepackage{comment}
\usepackage{breqn}
\usepackage{amsmath}
\usepackage{longtable}

\received{ }
\revised{ }
\accepted{ }
\submitjournal{ApJ}

\shorttitle{\Hnaut\ from Cluster-Lensed Quasars}
\shortauthors{Napier et al.}
\graphicspath{{./}{figures/}}

\begin{document}

\title{Hubble Constant Measurement from Three Large-Separation Quasars Strongly Lensed by Galaxy Clusters}

\correspondingauthor{Kate Napier}
\email{kanapier@umich.edu}

\author[0000-0003-4470-1696]{Kate Napier}
\affiliation{Department of Astronomy, University of
Michigan, 1085 S University Ave, Ann Arbor, MI 48109, USA}

\author[0000-0002-7559-0864]{Keren Sharon}
\affiliation{Department of Astronomy, University of
Michigan, 1085 S University Ave, Ann Arbor, MI 48109, USA}

\author[0000-0003-2200-5606]{H{\aa}kon Dahle}
\affiliation{Institute of Theoretical Astrophysics, University of Oslo, P.O. Box 1029, Blindern, NO-0315 Oslo, Norway}

\author[0000-0003-1074-4807]{Matthew Bayliss}
\affiliation{Department of Physics, University of Cincinnati, Cincinnati, OH 45221, USA}

\author[0000-0003-1370-5010]{Michael D. Gladders}
\affiliation{Department of Astronomy and Astrophysics, University of Chicago, 5640 South Ellis Avenue, Chicago, IL 60637, USA}

\author[0000-0003-3266-2001]{Guillaume Mahler}
\affiliation{Centre for Extragalactic Astronomy, Durham University, South Road, Durham DH1 3LE, UK}
\affiliation{
Institute for Computational Cosmology, Durham University, South Road, Durham DH1 3LE, UK}

\author[0000-0002-7627-6551]{Jane R. Rigby}
\affiliation{Observational Cosmology Lab, Code 665, NASA Goddard Space Flight Center, Greenbelt, MD 20771, USA}

\author[0000-0001-5097-6755]{Michael Florian}
\affiliation{Steward Observatory, University of Arizona, 933 North Cherry Ave., Tucson, AZ 85721, USA}







\begin{abstract}
Tension between cosmic microwave background-based and distance ladder-based determinations of the Hubble constant \Hnaut\ motivates pursuit of independent methods that are not subject to the same systematic effects. A promising alternative, proposed by Refsdal in 1964, relies on the inverse scaling of \Hnaut\ with the delay between the arrival times of at least two images of a strongly-lensed variable source such as a quasar. To date, Refsdal's method has mostly been applied to quasars lensed by individual galaxies rather than by galaxy clusters.  Using the three quasars strongly lensed by galaxy clusters (\SDSSfiveim, \SDSSthreeim, and \SDSSsixim) that have both multiband Hubble Space Telescope data and published time delay measurements, we derive \Hnaut, accounting for the systematic and statistical sources of uncertainty.  While a single time delay measurement does not yield a well-constrained \Hnaut\ value, analyzing the systems together tightens the constraint.  Combining the six time delays measured in the three cluster-lensed quasars gives \Hnaut\ = \Htot\ \unitsHnaut.  To reach 1$\%$ uncertainty in \Hnaut, we estimate that a sample size of order of 620 time delay measurements of similar quality as those from \SDSSfiveim, \SDSSthreeim, and \SDSSsixim\ would be needed.  Improving the lens modeling uncertainties by a factor of two and a half may reduce the needed sample size to 100 time delays, potentially reachable in the next decade.    
\end{abstract}

\keywords{galaxy clusters; quasars; time delay; Hubble constant}


\section{Introduction} \label{sec:intro}
The Hubble parameter \Hnaut, which describes the current expansion rate of the Universe, has been sought since the discovery in the 1920s that the Universe is expanding \citep{Lemaitre1927,Hubble1929}.  At the turn of the last century, measurements of \Hnaut\ started converging around \Hnaut\ = 70 \unitsHnaut.  However, as \Hnaut\ measurements have become increasingly precise, the so-called ‘Hubble Tension' has arisen between the estimates from early- and late-Universe probes.  The Planck Collaboration reported \Hnaut\ = 67.4 $\pm$ 0.5 \unitsHnaut\ \citep{Planck2020}.  They used density fluctuations encoded in the Cosmic Microwave Background (CMB) at the surface of last scattering to determine H at that epoch and then used a spatially flat cosmological model to extrapolate to \Hnaut.  By contrast, the ``Supernovae, \Hnaut, for the Equation of State of Dark Energy" (SH0ES) collaboration combined \textit{Gaia} parallaxes and multiband HST photometry of Milky Way Cepheids to calibrate the extragalactic distance scale and derive \Hnaut\ = 73.04 $\pm$ 1.04 \unitsHnaut\ \citep{Riess2022}.  The Planck and SH0ES values, which respectively capture the early and late-time physics of the Universe, differ by 5$\sigma$.  \cite{Freedman2021} applied an updated Tip of the Red Giant Branch (TRGB) calibration to a distant sample of Type Ia supernovae from the Carnegie Supernova Project and obtained \Hnaut\ = 69.8 $\pm$ 0.6 (stat) $\pm$ 1.6 (sys) \unitsHnaut, consistent with the CMB value, and within 2$\sigma$ of the SH0ES value, owing to the larger uncertainties.  The discrepancy between different \Hnaut\ methods may indicate a deviation from the standard $\Lambda$ Cold Dark Matter ($\Lambda$CDM) model, and therefore new physics, or the presence of unknown or underestimated systematics.  Either way, this tension remotivates the pursuit of other \Hnaut\ determination methods that are not prone to the same systematics.

An alternative \Hnaut\ determination method, proposed by \cite{Refsdal1964}, uses the fact that \Hnaut\ scales inversely with the delay between the arrival times of at least two images of a strongly-lensed variable source, such as a quasar or a supernova.  Due to the rarity of galaxy clusters lensing quasars or supernovae, the Refsdal \Hnaut\ technique has primarily been sought with galaxy-scale lenses \citep[see e.g., the recent reviews by][] {moresco2022, BirrerReview, Treu2022}.    

Of the $>$300 known lensed quasars, the vast majority are lensed by individual galaxies \citep{lemon19, lemon22}.  Quasars lensed by individual galaxies have been used to obtain \Hnaut.  For example, the \Hnaut\ Lenses in COSMOGRAIL's Wellspring (H0LiCOW) collaboration performed a joint analysis of six galaxy-lensed quasars to obtain \Hnaut\ = 73.3$^{+1.7}_{-1.8}$ \unitsHnaut\ \citep{Wong2020}.  This value is consistent with the Cepheid-calibrated measurement from the SH0ES collaboration.  

The main uncertainty with galaxy-scale lenses is the mass-sheet degeneracy which does not allow the total density profile to be precisely constrained from the lensing information alone.  \cite{Kochanek2020} argued that the H0LiCOW \Hnaut\ measurement uncertainty was underestimated due to artificially restricting the mass-sheet degeneracy through specific mass model assumptions.  To completely account for the mass-sheet degeneracy in the uncertainty, \cite{Birrer2020} permitted additional freedom in the lens model and constrained the mass profile with stellar kinematics.  Unsurprisingly, this led to an increase in the uncertainty level to $\sim$8$\%$, but the mean \Hnaut\ value remained at $\sim$73 \unitsHnaut.  To reduce the uncertainty, \cite{Birrer2020} included further information on the galaxy mass profile from the Sloan Lens ACS sample and obtained \Hnaut\ = 67.4 $\pm$ 3.7 \unitsHnaut.  Resolved JWST and Extremely Large Telescope observations of the stellar kinematics in the lens galaxies may significantly reduce these sources of systematic errors \citep{Birrer2021}.

What has remained largely unexplored until now is determining \Hnaut\ by using quasars that are strongly lensed by galaxy clusters.   For several reasons, cluster-lensed quasars can potentially overcome some of the difficulties faced by individual galaxy lenses.  First, since galaxy clusters have deeper potential wells than galaxies, cluster lenses produce longer time delays of order months to years compared to typically a month in galaxy lenses.  Consequently, the observationally measured time delay values will have smaller fractional uncertainty, which then will propagate to reduced uncertainty in \Hnaut\ due to the inverse scaling of \Hnaut\ with time delays.  Second, the light curves of cluster-lensed quasars are less likely to be affected by microlensing from stars in the lens plane, because the mass distribution is dominated by dark matter at the projected radius at which the images appear.  Third, galaxy cluster mass distributions are less affected by complex baryonic physics than those of galaxy lenses; the complex baryonic surface density of galaxy-scale lenses may be a significant source of systematic uncertainty. 
 Cluster-scale lenses are less affected by the mass-sheet degeneracy due to having multiple images of lensed sources from various source plane redshifts behind the lens \citep{Grillo2020}.
A challenge that must be contended with, however, is the complexity of cluster lenses.

Two inputs are necessary to use cluster-lensed quasars to determine \Hnaut.  The first is an observational measurement of the time delay between the multiple quasar images, and the second is an accurate mapping of the projected density of the dark and luminous mass at the cluster core.  High accuracy lens models require space-based resolution and spectroscopic follow-up.  Of the seven published cluster-lensed quasars to date \citep{Inada2003, Inada2006, Dahle2013, Shu2018, Shu2019, Martinez2022, Napier2023}, only three have the necessary data available to determine \Hnaut: \SDSSfiveim, \SDSSthreeim, and \SDSSsixim.  In this paper, we use the available archival HST data and the latest measurements of time delay and spectroscopic redshifts of background sources from the literature to obtain an independent measurement of \Hnaut\ from these three systems.

This paper is organized as follows: In Section \ref{sec:analysis}, we outline the theory of observational gravitational lensing time delay and its dependence on \Hnaut.  In Section \ref{sec:lm} we detail the lens modeling procedure.   In Sections \ref{sec:1004}, \ref{sec:1029}, and \ref{sec:2222}, we give an overview of the three cluster-lensed quasar systems used in this \Hnaut\ analysis and provide details about their \HST\ and spectroscopic data, time delays, and lens models.  In Section \ref{sec:results}, we present our constraints on \Hnaut.  We conclude in Section \ref{sec:discussion} with a discussion of our \Hnaut\ result and the future prospects of the time delay \Hnaut\ method.

Throughout the paper, we adopt the standard $\Lambda$CDM flat cosmological model with $\Omega_{m}$ = 0.3 and $\Omega_{\Lambda}$ = 0.7.

\section{Time Delay Analysis}\label{sec:analysis}
The Refsdal \Hnaut\ method is possible due to the measurable delay between the arrival time of two or more images of a variable source such as a quasar.  Under the thin lens approximation, a packet of light that travels from the source to the observer will be delayed by time \textit{t} given by the arrival time surface \citep{Schneider1985}:   

\begin{equation}\label{arrival_time}
t(\vec{\theta}, \vec\beta) = \frac{1+z_l}{c}\frac{d_l d_s}{d_{ls}}[ \frac{1}{2} (\vec\theta-\vec\beta)^2-\psi(\vec\theta)],
\end{equation}
\vspace{-0.2in}

\noindent where $z_l$ is the redshift of the lens, $d_l$, $d_s$, and $d_{ls}$ are the angular diameter distances from the observer to the lens, to the source, and between the lens and the source, respectively; $\vec\theta$ is the image position in the image plane; $\vec\beta$ is the unobserved source position; and $\psi(\vec\theta)$ is the gravitational lensing potential.  The arrival time \textit{t} is a combination of the path length and the gravitational time delay (\textit{t} = \textit{t$_{geometric}$} + \textit{t$_{grav}$}). The last term, $\tau(\theta;\beta)=[ \frac{1}{2} (\vec\theta-\vec\beta)^2-\psi(\vec\theta)]$, is also known as the Fermat potential. The multiple images of a strongly-lensed source appear in the stationary points of the arrival time surface, that is, in the minima, maxima, and saddle points.  \Hnaut\ is incorporated in Eq.~\ref{arrival_time} because of its inverse scaling with the angular diameter distances:   
\begin{equation}\label{ad_dist}
d_A(z_1,z_2)=\frac{1}{(1+z_2)}\frac{c}{H_0}\int\limits_{z_1}^{z_2}{\frac{dz}{E(z; \Omega_m, \Omega_{\Lambda})}},
\end{equation}

\noindent for flat cosmology, where $E(z; \Omega_m, \Omega_{\Lambda})$ is a dimensionless function given by $E(z; \Omega_m, \Omega_{\Lambda})=\sqrt{\Omega_m(1+z)^3+\Omega_{\Lambda}}$.  The matter density and vacuum energy density parameters are $\Omega_m$ and $\Omega_{\Lambda}$, respectively.  Conveniently, \Hnaut\ is disentangled from the other cosmological parameters in the angular diameter distance equation (Eq.~\ref{ad_dist}).  After substituting Eq.~\ref{ad_dist} into $d_ld_s/d_{ls}$ in Eq.~\ref{arrival_time}, the time delay is determined by applying Eq.~\ref{arrival_time}\ for two image positions corresponding to the same source position and taking the difference.  The time delay between the images thus becomes: 

\begin{equation}\label{time_delay}
\begin{split}
\Delta t &= \left( \frac{1}{H_0}\right) \left( \frac{\int\limits_{0}^{z_l}{\frac{dz}{E(z)}}  \int\limits_{0}^{z_s}{\frac{dz}{E(z)}}}{\int\limits_{z_l}^{z_s}{\frac{dz}{E(z)}}}\right) \times \\
&\left(\frac{1}{2} [(\vec\theta_1-\vec\beta)^2 - (\vec\theta_2-\vec\beta)^2 ]-[\psi(\vec\theta_1)-\psi(\vec\theta_2)]\right)
\vspace{-0.1in}
\end{split}
\end{equation}

The first term on the right-hand side of the time delay equation gives the Hubble parameter; the second term contains the dependence on cosmological parameters other than \Hnaut; and the third term is solved by the strong gravitational lens model.  We neglect the higher order effects of the cosmological parameters and fix the second term on the right-hand side in Eq. \ref{time_delay} using the fiducial cosmology.  The left-hand side of the equation is the measurement of the time delay, e.g., from monitoring and comparing the observed light curves of two images of the variable source.  

Once we compute a model of the lensing mass distribution (see Section~\ref{sec:lm}), we determine the model-predicted excess arrival time surface (Eq.~\ref{time_delay}) with respect to one of the quasar images.  Assuming our lens model is a correct description of the matter distribution, we then leverage the fact that the time delay scales inversely with \Hnaut.  We compare the model-predicted time delays between images to the observational measurements of the time delays to obtain \Hnaut\ via:
\begin{equation}\label{getH0}
H_{0} = H_{0,model} \times \frac{\Delta t_{model}}{\Delta t_{measured}}
\end{equation} 
where H$_{0,model}$ is the \Hnaut\ value used to generate the Fermat potential from the lensing analysis (we assumed H$_{0,model}$ = 70 \unitsHnaut), $\Delta$t$_{model}$ is the model-predicted time delay between the quasar images, and $\Delta$t$_{measured}$ is the observational measurement of the time delay between the pair of quasar images.

\section{Lens modeling}\label{sec:lm}
We computed the lens models with the publicly available software \Lenstool\ \citep{Jullo2007}.  \Lenstool\ is a ‘parametric' modeling algorithm which describes the lensing mass distribution as a linear combination of galaxy-scale, group-scale, and cluster-scale halos, each of which is parameterized as a pseudo-isothermal ellipsoidal mass distribution \citep[PIEMD, also called dPIE;][]{Eliasdottir2007}.  
The three-dimensional density distribution of the PIEMD,
\begin{equation}\label{piemd}
\rho(r) = \frac{\rho_{0}}{(1+r^{2}/r^{2}_{c})(1+r^{2}/r^{2}_{cut})},
\end{equation} has seven parameters whose values can either be fixed or varied: position (\textit{x}, \textit{y}); ellipticity \textit{e} = ($\textit{a}^{2}$-$\textit{b}^{2}$)/($\textit{a}^{2}$+$\textit{b}^{2}$), where \textit{a} and \textit{b} are the semi-major and semi-minor axes, respectively; position angle \textit{$\theta$}; core radius \textit{r$_{c}$}; truncation radius \textit{r$_{cut}$}; and effective velocity dispersion \textit{$\sigma_{0}$}.  The parameters of the group-scale and cluster-scale halos are typically allowed to vary.  The exception is \textit{r$_{cut}$} for the cluster-scale halos as this radius usually occurs outside the region where strong lensing evidence is found, and thus, cannot be constrained.  

\Lenstool\ uses a Markov Chain Monte Carlo (MCMC) sampling of parameter space.  The best-fit model is identified as the one that minimizes the scatter between the model-predicted and observed image locations in the image plane (``image plane minimization'') or minimizes the scatter between the predicted source locations of multiple images in the source plane (``source plane minimization'').  The lens models employ the strong lensing information of multiply-imaged galaxies (arcs), whose positions and redshifts are used as model constraints.  The availability of lensing constraints strongly affects the accuracy of lens models, as they are used as local solutions of the lensing equations and constrain the projected mass density distribution at the cluster's core.  The mass distribution and magnification are sensitive to the accurate identifications and positions of multiple images and to the redshifts of the lensed galaxies.  It is necessary to include a few spectroscopic redshifts in the lens model to avoid incorrect results \citep{JohnsonSharon2016}.  As described in more detail in the following sections, all of the multiply-imaged galaxies used as constraints in the lens models for \SDSSfiveim, \SDSSthreeim, and \SDSSsixim\ were robustly determined; hence the image identifications are not a significant source of systematic uncertainty.

To select cluster-member galaxies, we followed the procedure of \cite{GladdersYee2000}, by selecting galaxies that fall on the cluster red sequence in a color-magnitude diagram. For \SDSSthreeim\ we also incorporated spectroscopic redshift information (see Section~\ref{sec:1029}). The galaxy-scale halos' positional parameters (\textit{x}, \textit{y}, \textit{e}, \textit{$\theta$}) are measured with \texttt{Source Extractor} \citep{BertinArnouts1996} and fixed.  The\textit{r$_{c}$}, \textit{r$_{cut}$}, and \textit{$\sigma_{0}$} of the galaxy-scale halos are scaled to their observed luminosity following the scaling relations in \cite{Limousin2005}.  A potential misidentification of cluster member galaxies is unlikely to be a significant source of systematic uncertainty.  For example, \cite{Napier2023} found that excluding a subsample of cluster members from the lens model did not significantly change the predicted time delays.   

\section{\SDSSfiveim}\label{sec:1004}   
\SDSSfiveim\ was the first discovered galaxy cluster strongly lensing a quasar \citep{Inada2003}.  The cluster at \z\ = 0.68 strongly lenses a quasar at \z\ = 1.734 into five images, with a maximum image separation of $14\farcs6$ (Table \ref{table:quasar_table}).  The cluster also strongly lenses several background sources at \z\ = 2.74 \citep{Sharon2005}, \z\ = 3.288 \citep{Keren_thesis, Oguri2010}, and \z\ = 3.332 \citep{Sharon2005} (Table~\ref{table:constraints}). 

We used archival HST multi-color imaging from the \textit{Advanced Camera for Surveys} (ACS).  The \SDSSfiveim\ imaging data include GO-10509 (PI: Kochanek) ACS/F814W, F555W, F435W (10 orbits); GO-9744 (PI: Kochanek) ACS/F814W, F555W (2 orbits); and GO-10793 (PI: Gal-Yam) ACS/F814W (1 orbit).  These data were originally proposed to identify multiply-imaged galaxies to construct a mass model \citep{Sharon2005}, search for the fifth quasar image \citep{Inada2005}, derive $\Omega_{\Lambda}$, perform a weak lensing analysis, and search for supernovae in massive high-redshift clusters \citep{Sharon2010}.  These data also enabled studies of the spectral energy distribution of the quasar host galaxy \citep{Ross2009}, the ultraviolet upturn in red sequence galaxies \citep{Ali2018}, and active galactic nuclei in massive clusters \citep{Klesman2012}.   

We modeled \SDSSfiveim\ using one cluster-scale halo, one brightest cluster galaxy (BCG)-scale halo, and a galaxy-scale halo for each of the cluster member galaxies, 168 in total, five (including the BCG) of which have their slope parameters optimized instead of adopting the scaling relations from \cite{Limousin2005}.  We optimized the parameters for these five galaxies because of either their proximity to the quasar images or their necessity for reproducing the lensed image configuration. 

We modeled the cluster using both source-plane minimization and image-plane minimization, and evaluated the quality of the models obtained by each approach. While formally the image-plane minimization resulted in a better image-plane scatter, these models  produced additional quasar images that are not observed.  Therefore, we proceeded with the source-plane minimization for \SDSSfiveim\ for the remainder of the analysis.  We note that the best-fit lens model produced large scatter between the observed and model-predicted positions in the image plane for quasar image C.  In our results, we checked what happens when image C is removed from the \Hnaut\ measurement.

\begin{figure*}
    \centering
    \includegraphics[width=500px]{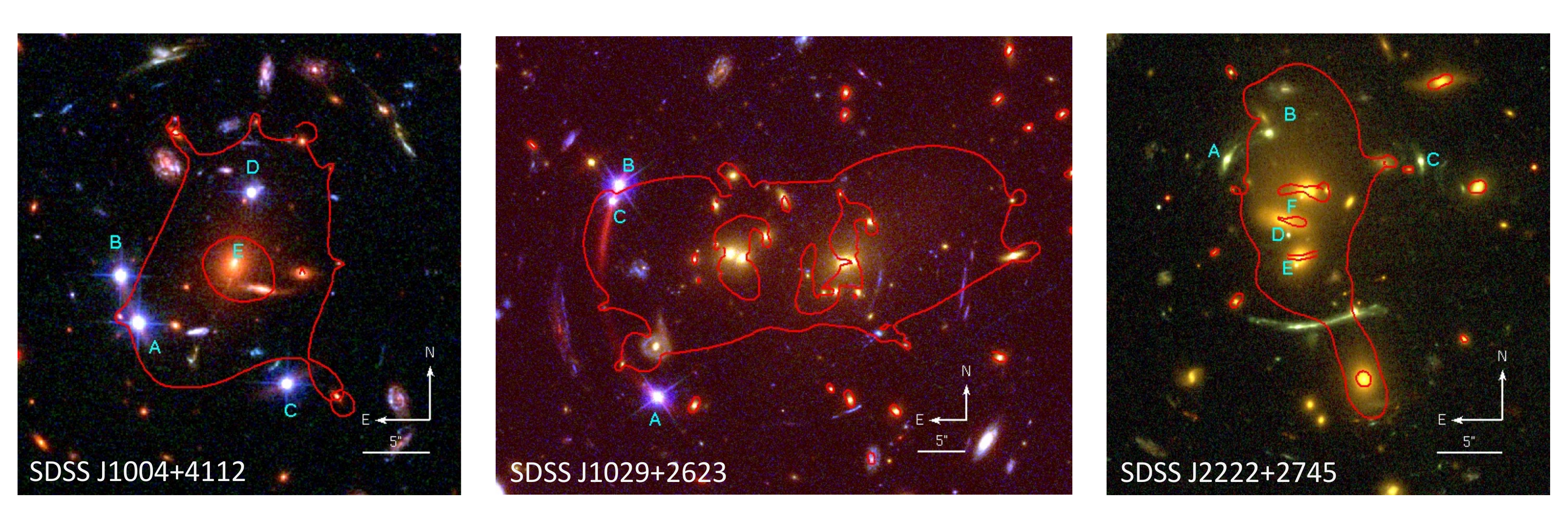}
    \caption{\fullHST\ imaging of the three cluster-lensed quasars used to derive \Hnaut.  We computed the lens models for \SDSSfiveim\ and \SDSSthreeim.  \SDSSsixim\ is reproduced from \citet{Sharon2017}.  The positions of the quasar images are denoted with the cyan letters.  The critical curves, the loci of maximum magnification at a specified source redshift, are generated at the quasar redshifts -- \textit{z} = 1.734, \textit{z} = 2.1992, and \textit{z} = 2.805, for \SDSSfiveim, \SDSSthreeim, and \SDSSsixim, respectively, and are plotted in red.}
    \label{fig:lens_models}
\end{figure*}

The model consists of \numParamsSDSSfiveim\ free parameters and \numConSDSSfiveim\ constraints.  The \HST\ data and the lens model for \SDSSfiveim\ are shown in Figure \ref{fig:lens_models}.  The redshifts of the arcs in our lens model are the same as those used by \cite{fores2022}.  The strong lensing mass model parameters are reported in Table~\ref{table:1004_model}.

The measured time delay between images A and B ($\Delta$t$_{AB}$ = -38.4 $\pm$ 2.0 days) was first published in \cite{Fohlmeister2007}.  In this notation, a positive value of the time delay means image A leads the other image.  In addition to reporting a refined value of $\Delta$t$_{AB}$ = -40.6 $\pm$ 1.8 days, \cite{Fohlmeister2008} measured the time delay between images A and C ($\Delta$t$_{AC}$ = -821.6 $\pm$ 2.1 days) and set a lower limit of $\Delta$t$_{AD}$ $>$ 1250 days.   After the completion of a 14.5 year monitoring campaign at the 1.2m Fred Lawrence Whipple Observatory (FLWO), \cite{Munoz2022} presented new light curves for the four brightest images in \SDSSfiveim, resulting in updated time delay values of $\Delta$t$_{AB}$ = -43.01 $\pm$ 0.27, $\Delta$t$_{AC}$ = -825.23 $\pm$ 0.46 days, and $\Delta$t$_{AD}$ = 1633.23 $\pm$ 0.97 days (Table~\ref{table:time_delays}).

\section{\SDSSthreeim}\label{sec:1029}
\SDSSthreeim\ is a cluster at \z\ = 0.588 that is strongly lensing a quasar at \z\ = 2.1992 into three images \citep{Inada2006, Oguri2008}.  The quasar images are in a naked cusp configuration with a maximum image separation of $22\farcs5$ (Table \ref{table:quasar_table}).    

\cite{Acebron2022} reported spectroscopic redshifts of several galaxies in the field, based on Multi Unit Spectroscopic Explorer (MUSE) spectroscopy from the Very Large Telescope.  They refined the redshift measurement of the quasar to \z\ = 2.1992 (formerly reported as \z\ = 2.197, \cite{Inada2006}).  The other spectroscopically confirmed objects from MUSE include a  doubly-imaged galaxy at \z=2.1812, a septuply-imaged galaxy at \z=3.0275, a quadruply-imaged galaxy at \z=3.0278, a doubly-imaged galaxy at \z=1.0232, and a quadruply-imaged galaxy at \z=5.0622 \citep{Acebron2022} (Table \ref{table:constraints}).  

We used archival HST multi-color imaging from GO-12195 (PI: Oguri): WFC3/F160W (2 orbits), ACS/F814W (3 orbits), and ACS/F475W (2 orbits).  These data were originally proposed to identify multiply-imaged galaxies to construct a mass model that could be used to better understand the anomalous flux ratios between two of the quasar images and the dynamical state of the cluster \citep{Oguri2013}.  These \HST\ data also enabled a weak lensing analysis and a morphology study of the quasar host galaxy \citep{Oguri2013}.  

Our lens model, which builds on the results from \cite{Acebron2022} and \cite{Oguri2013}, contains \numConSDSSthreeim\ constraints and \numParamsSDSSthreeim\ free parameters.  All of the model constraints are taken from \cite{Acebron2022}.  The model includes two cluster-scale dark matter halos that were allowed to vary in position around the two BCGs as well as two galaxy-scale halos that were fixed to the BCGs' positions.  Additionally, a foreground galaxy (\z=0.5111 from MUSE) and a background galaxy (\z=0.6735 from MUSE) along the line of sight are both modeled at the cluster redshift since \Lenstool\ does not yet implement a multi-plane lensing framework.  This approach improves the accuracy of the lensing analysis outputs compared to omitting these interlopers from the model \citep{Raney2020}.  The lens model includes 204 galaxy-scale halos.  

Our lens model differs from \cite{Acebron2022} in the following ways.  Whereas \cite{Acebron2022} include a model (Model 1) with an external shear component, we opted to not include this component as its inclusion does not significantly improve the rms scatter in the source plane.  Additionally, for consistency with the other clusters modeled in this paper, our galaxy-scale halos have ellipticities, whereas \cite{Acebron2022} use spherical halos.  We constructed our galaxy catalog as described in Section \ref{sec:lm}, taking into account the MUSE spectroscopy to determine the red sequence \citep[see][]{Sharon2022b}.  We used the ACS F814W vs. F475W for selection.  We identified the red sequence by fitting a line to the spectroscopic members in this phase space, with four iterations of sigma clipping.  

We found that the source-plane minimization did a better job at predicting the quasar image positions in this cluster than the image-plane minimization, possibly due to the close proximity of quasar images B and C. 
Once a best-fit model was obtained, we examined the posterior distribution of image predictions and rejected from the MCMC sampling steps that did not produce this lensing configuration, i.e., not producing two separate images for A and B on either side of the critical curve. Since these two images lie very close to the critical curve, some parameter combinations produce solutions in which these two images merge and only image A of the quasar remains, in contrast to the observed lensing evidence.

The \HST\ data and the lens model for \SDSSthreeim\ are shown in Figure \ref{fig:lens_models}.  The strong lensing mass model parameters are reported in Table \ref{table:1029_model}.   

\cite{Fohlmeister2013} published 
the time delay measurement between images A and B ($\Delta$t$_{AB}$ = 744 $\pm$ 10 days) based on photometric monitoring campaign at the FLWO 1.2m.   

\section{\SDSSsixim}\label{sec:2222}

\SDSSsixim, discovered by \cite{Dahle2013}, is a cluster at \textit{z} = 0.49 that strongly lenses a quasar at z = 2.805.  The quasar is imaged six times \citep{Sharon2017} with a maximum image separation of $15\farcs1$ (Table \ref{table:quasar_table}). 

Spectroscopy of other lensed galaxies was obtained by the Gemini North Telescope.  These data include triply-imaged and doubly-imaged knots from a galaxy at \z\ = 4.562 and a doubly-imaged galaxy at \z\ =  2.2963 \citep{Sharon2017}.  

We used archival HST multi-color imaging from GO-13337 (PI: Sharon): WFC3/F160W, F110W (1 orbit) and ACS/F814W, F606W, F435W (6 orbits).  These data were originally proposed to detect any additional quasar images 
and to compute a mass model \citep{Sharon2017}.  Additionally, these \HST\ data have enabled a spatially resolved study of the Lyman-alpha emission in the quasar host galaxy \citep{Bayliss2017}.  

We adopted the lens model from \cite{Sharon2017} with \numConSDSSsixim\ constraints and \numParamsSDSSsixim\ free parameters.  \SDSSsixim\ is modeled using image plane minimization with one cluster-scale halo and 172 galaxy-scale halos. \cite{Sharon2017} included as constraints triply-imaged and doubly-imaged knots at the quasar's redshift of $\textit{z}$ = 2.805, and triply-imaged and doubly-imaged knots from a galaxy at $z$ = 4.562.  Two separate triply-imaged galaxies had their redshifts left as free parameters, with priors of 2.0 $\le$ \textit{z} $\le$ 4.0  and 3.8 $\le$ \textit{z} $\le$ 5.0, respectively, based on photometric redshift analysis.  The \HST\ data and the lens model for \SDSSsixim\ are shown in Figure \ref{fig:lens_models}.  Table 5 of \cite{Sharon2017} lists the strong lensing mass model parameters. 

\cite{Dahle2015} first published the time delay measurements between images A and B ($\Delta$t$_{AB}$ = 47.7 $\pm$ 6.0 days) and A and C  ($\Delta$t$_{AC}$ = -722 $\pm$ 24 days).  Then \citet{Dyrland2019} reported updated values for the time delays between images A and B ($\Delta$t$_{AB}$ = 42.44 $^{+1.36}_{-1.44}$ days) and images A and C ($\Delta$t$_{AC}$ = -696.65 $^{+2.00}_{-2.10}$ days).  These measurements were based on data from a monitoring campaign at the 2.5m Nordic Optical Telescope.  

In the analysis that follows, we used the most up-to-date time delay values for \SDSSfiveim, \SDSSthreeim, and \SDSSsixim\ which are listed in Table \ref{table:time_delays}.

\begin{figure}[h]
    \centering
    \includegraphics[width=260px]{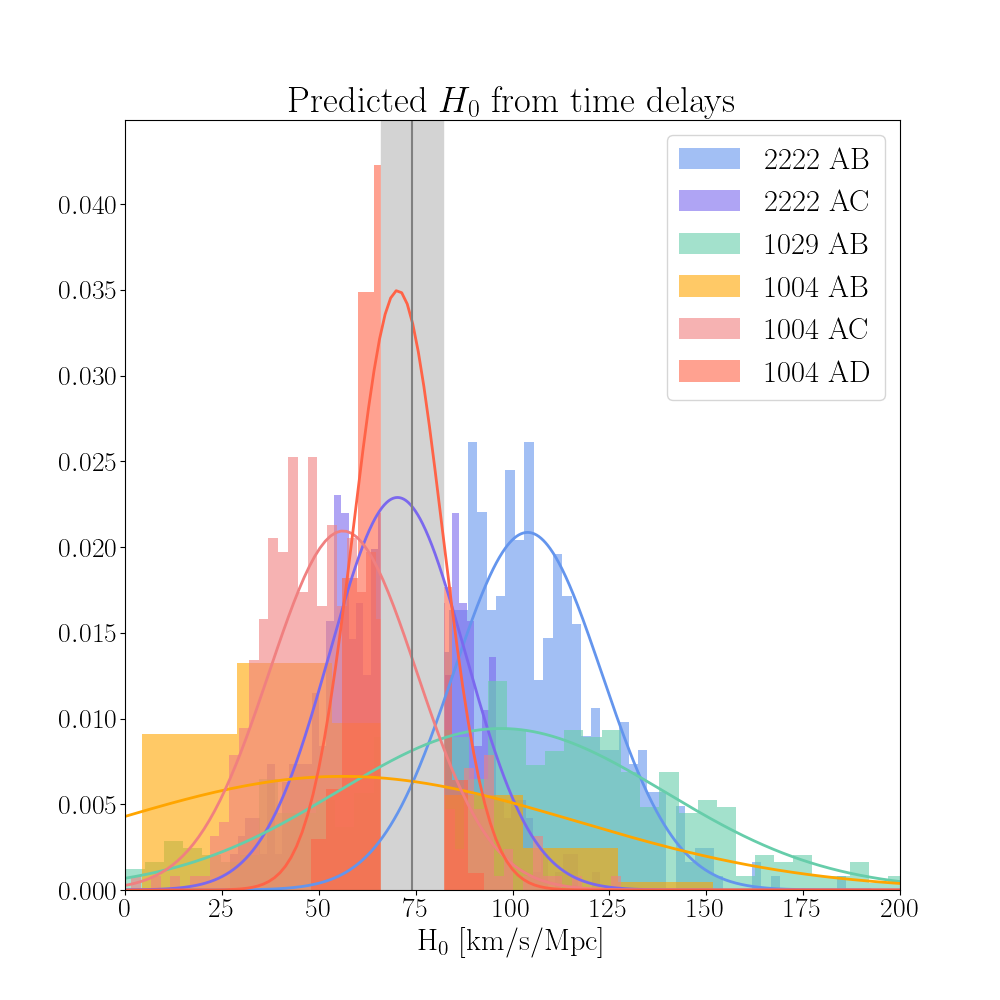}
    \caption{Constraints on \Hnaut\ from three cluster-lensed quasars, \SDSSfiveim, \SDSSthreeim, and \SDSSsixim.  The histograms are created from 500 random models sampled from the MCMC. Overplotted are Gaussian fits to the distributions. Whereas individual time delay measurements produce \Hnaut\ values with an average of 41$\%$ error, the error is decreased to 11$\%$ when the systems are analyzed together.  The inverse-variance weighted mean of \Hnaut\ is \Htotvalue\ \unitsHnaut\ (solid gray line) and the standard error of the weighted mean is \Htoterr\ \unitsHnaut.}
    \label{fig:Hnaut}
\end{figure}
\vspace{0.75in}
\section{Results} \label{sec:results}
Using the outputs of the lens models described in the previous sections, 
we computed the model-predicted time delay values for each of the quasar images in each cluster field with respect to image A of the quasar (Equation \ref{time_delay} and Table \ref{table:pred_time_delays}).

The time delay is a sensitive function of the positions of the source ($\vec{\beta}$) and its multiple images ($\vec{\theta_1}$,$\vec{\theta_2}$). The unobservable source position and the locations of its multiple images are strongly coupled to the time delay, since stationary points in the arrival time surface determine the image-plane positions of multiple images of any given source-plane position (see Section~\ref{sec:analysis}). It is therefore important to measure time delays self-consistently, by obtaining the time delay at the image positions predicted by the same lensing potential. Lens models are never perfect, and small scatter between the observed and predicted positions is expected. 
To maintain this self-consistency, we calculated the source position $\vec{\beta}$ by ray-tracing the observed position of image A ($\vec{\theta}_{\rm A}$) through the lens equation, and used the same lens model to predict the image-plane positions of its counter images ($\vec{\theta_2}$,$\vec{\theta_3}$,...). The time delay was then calculated from these predicted positions, rather than the observed positions, which may be slightly shifted from the stationary points in the Fermat potential. The scatter in the image or source plane contributes to the error budget through the MCMC exploration of the parameter space. An alternative approach to determining the source position is averaging the predicted source locations from all the quasar images, and calculating the predicted image locations of the average source.  We found that this alternative approach of determining the source position produced an \Hnaut\ value consistent with the method described above, and hence, the choice of one approach vs. the other is not a significant source of systematic uncertainty.  In what follows, we describe the results using the former method.

Using Equation \ref{getH0}, we computed the \Hnaut\ value corresponding to each independent published time delay value and corresponding predicted time delays.
To generate the 1$\sigma$ uncertainties in \Hnaut, we used 500 random models from the MCMC sampling of the parameter space for each cluster to ensure smooth sampling of the \Hnaut\ posteriors. 

The number of measured time delays in each field determines the number of \Hnaut\ measurements derived from each cluster: three from \SDSSfiveim, one from \SDSSthreeim, and two from \SDSSsixim, for a total of six \Hnaut\ measurements.  Table \ref{table:Hnaut_constraints} lists the derived \Hnaut\ values and uncertainties, obtained for the ``best'' lens model, i.e., the one producing the smallest scatter, and for the full posterior distribution.  We find that the fractional uncertainty on \Hnaut\ calculated using the lensing analysis is consistent with the analytical propagation of errors due to the source plane uncertainty, following \citet{BirrerTreu2019} Eq. 16 and the examples therein, giving $\delta H_{0} (\delta \theta)/H_{0} = 15-50\%$. 

The resulting \Hnaut\ measurement from each quasar pair has large uncertainties due to the complexity of the lens and systematic uncertainties in the lens modeling process.  \cite{Grillo2020} found that systematic effects such as adding a constant mass sheet at the cluster redshift and using a power-law profile for the mass density of the cluster's primary halo did not introduce a significant bias on the inferred \Hnaut\ value and that the statistical uncertainties dominate the total error budget.  Given that \SDSSfiveim, \SDSSthreeim, and \SDSSsixim reside in the same universe, they all must have the same \Hnaut; we can leverage these three independent lines of sight, with six time delays, to obtain a tighter constraint than what is possible from a single time delay, and marginalize over systematics related to line of sight effects such as intervening structure.  We combine the results from the six time delays by taking the inverse-variance weighted mean of the six \Hnaut\ measurements, sampled from their posterior distributions. We accounted for the correlations between measurements made in the same cluster lens by using the same model realization from the MCMC to generate a linked set of \Hnaut\ values.  For example, since \SDSSfiveim\ has three measured time delays, the same model realization was used to calculate the \Hnaut\ values from $\Delta t_{AB}$, $\Delta t_{AC}$, and $\Delta t_{AD}.$ 

We note that the observational time delay measurement uncertainties are negligible compared to the lens modeling uncertainties.  The inverse-variance weighted mean and the standard error of the weighted mean of \Hnaut\ is \Htot\ \unitsHnaut\ (Fig.~\ref{fig:Hnaut}). Combining the \Hnaut\ values derived from multiple time delay values improves the constraints on \Hnaut, decreasing the uncertainty from $\sim$41$\%$ for an individual \Hnaut\ measurement to 11$\%$ for the sample.  If \SDSSfiveim's  quasar image C is excluded from the analysis (see Section~\ref{sec:1004}), we obtain \Hnaut\ = 77.4 $\pm$ 9.9 \unitsHnaut.  This value is consistent with the \Hnaut\ measurement calculated with all six time delays.

\section{Discussion}
\label{sec:discussion}
Our analysis provides an independent \Hnaut\ measurement that is not sensitive to the same systematics as other methods.  Our \Hnaut\ measurement, \Htot\ \unitsHnaut (or \Hnaut\ = 77.4 $\pm$ 9.9 \unitsHnaut\ if excluding \SDSSfiveim\ AC),  is higher than the values from CMB (67.4 $\pm$ 0.5 \unitsHnaut; \citealt{Planck2020}), TRGB (69.8 $\pm$ 0.6 (stat) $\pm$ 1.6 (sys); \citealt{Freedman2021}), and Cepheids (73.04 $\pm$ 1.04 \unitsHnaut; \citealt{Riess2022}).  However, given our measurement's larger fractional uncertainty, our \Hnaut\ value is consistent with all three.

Increasing the number of systems used for a combined time-delay measurement of \Hnaut\ will improve this method's competitiveness with CMB-based and distance ladder-based methods.  Although four other cluster-lensed quasars are published in the literature, none has all the necessary time delays measurements, space-resolution imaging, and spectroscopic redshifts of secondary arcs for a measurement of \Hnaut.  All four of the other published cluster-lensed quasars have ongoing photometric monitoring campaigns to measure their time delays.  Additionally, one of the other four systems, COOL J0542-2125 \citep{Martinez2022} will be observed by \HST\ in Cycle 30 (GO-17243; PI: Napier).

To estimate the improvement in the \Hnaut\ constraint from a sample of twice as many time delay measurements, we simulated \Hnaut\ distributions guided by the existing sample, as follows.  We sampled integer \Hnaut\ values from a Gaussian distribution centered at our combined \Hnaut\ value of 74.1 \unitsHnaut.  We then randomly assigned to these six \Hnaut\ values the standard deviation of one of the six \Hnaut\ distributions (Table \ref{table:Hnaut_constraints}), and produced the corresponding Gaussian distributions.  We repeated this simulation process 100,000 times. Incorporating these new six \Hnaut\ distributions for a total of 12 constraints, and averaging the 100,000 iterations, gave a standard error of the weighted mean of 5.5 \unitsHnaut.  Therefore, doubling the number of systems results in a $\sim$35-40$\%$ improvement in the constraint on \Hnaut, reducing the uncertainty on \Hnaut\ from 11\% to $\sim$7.4\%. 

A 1$\%$ uncertainty measurement of \Hnaut\ from cluster-lensed quasars would be competitive with the current precision level of CMB and distance ladder methods.  Extending the simulation described above to a larger number of systems, we estimated that 620 delay measurements from cluster-lensed quasars would achieve a 1$\%$ uncertainty level on \Hnaut\ from cluster lensed-quasars.  Based on \SDSSfiveim, \SDSSthreeim, and \SDSSsixim\ each having an average of two time delay measurements, a sample size of 310 cluster-lensed quasars would be needed to produce 620 time delay measurements.  Future surveys are expected to detect of order $\sim$50 such systems in the next decade \citep{Robertson2020}.  Therefore, this increase in sample size alone will not achieve 1$\%$ uncertainty in \Hnaut;  to reach 1$\%$ with of order of 50 systems (100 time delays) will require a decrease in the lens modeling uncertainties by about a factor of two and a half, on average.  Future work will explore whether this decrease in the uncertainties is feasible.  

\acknowledgments

%

Based on observations made with the NASA/ESA {\it Hubble Space Telescope}, obtained from the Multimission Archive at the Space Telescope Science Institute (MAST) at the Space Telescope Science  Institute, which is operated by the Association of Universities for Research in Astronomy, Incorporated, under NASA contract NAS 5-26555. These archival observations are associated with programs GO-10509, GO-9744, GO-10793, GO-12195, and GO-13337. Support for \HST\ program AR-16150, which enabled this work, was provided through grants from the STScI under NASA contract NAS5-26555.  Co-author GM received funding from the European Union's Horizon 2020 research and innovation programme under the Marie Sklodowska-Curie grant agreement No 896778.  We thank Ana Acebron for her useful discussions about \SDSSthreeim.  We thank the anonymous referee for their constructive feedback and comments that improved the manuscript.

\facilities{HST(ACS); HST(WFC3); HST(MAST)}

\software{\texttt{Lenstool} \citep{Jullo2007}; \texttt{Source Extractor} \citep{BertinArnouts1996}}




\clearpage
\bibliography{bibliography.bib}
\bibliographystyle{aasjournal}

\begin{table}[h!]
\begin{center}
\begin{tabular}{c c c c c c c}
 \hline
 Target & QSO Image & QSO \textit{z} & RA [J2000] & Decl. [J2000] & $\mu$ \\
 \hline
  SDSS J1004+4112 & & & & & & \\
  & A & 1.734 & 151.1450074 & 41.2109193 & 23.3$\pm$7.2  \\
  & B & 1.734 & 151.1454888 & 41.2119003 & 13.8$\pm$3.8 \\
  & C & 1.734 & 151.1409266 & 41.2096668 & 10.7$\pm$1.3 \\
  & D & 1.734 & 151.1419060 & 41.2136092 & 4.7$\pm$1.1 \\
  & E & 1.734 & 151.1423413 & 41.2122017 & 0.2$\pm$0.05 \\
  SDSS J1029+2623 & & & & & & \\
  & A & 2.1992 & 157.3081009 & 26.3883044 & 5.5$\pm$0.3 \\
  & B & 2.1992 & 157.3093619 & 26.3944624 & 20.8$\pm$3.9 \\
  & C & 2.1992 & 157.3095755 & 26.3939894 & 8.0$\pm$8.0 \\
  SDSS J2222+2745 & & & & & & \\
  & A & 2.805 & 335.537707 & 27.760543 & 12.0$\pm$5.0 \\
  & B & 2.805 & 335.536690 & 27.761119 & 8.4$\pm$5.0 \\
  & C & 2.805 & 335.532960 & 27.760505 & 5.5$\pm$2.0 \\
  & D & 2.805 & 335.536205 & 27.758901 & 0.9$\pm$0.5 \\
  & E & 2.805 & 335.536007 & 27.758248 & 0.6$\pm$0.2 \\
  & F & 2.805 & 335.535874 & 27.759723 & 0.7$\pm$0.4 \\
\end{tabular}
\caption{The quasar image positions and redshifts.  Also included are the model-predicted median magnifications at the observed positions of the quasar images.}
\label{table:quasar_table}
\end{center}
\end{table}

\begin{center}
\begin{longtable}{lllll}
\hline
System & ID & R.A. [J2000] & Decl. [J2000] & \textit{z} \\
\hline
\SDSSfiveim & & & & \\
& QSO-A & 151.1450074 & 41.2109193 & 1.734 \\
& QSO-B & 151.1454888 & 41.2119003 & 1.734 \\
& QSO-C & 151.1409266 & 41.2096668 & 1.734 \\
& QSO-D & 151.1419060 & 41.2136092 & 1.734 \\
& QSO-E & 151.1423413 & 41.2122017 & 1.734 \\
& 2.1 & 151.1418821 & 41.2102917 & 2.74 \\
& 2.2 & 151.1468800 & 41.2153908 & 2.74 \\ 
& 21.1 & 151.1417325 & 41.2103272 & 2.74 \\
& 21.2 & 151.1470383 & 41.2153011 & 2.74 \\
& 21.3 & 151.1419526 & 41.2116044 & 2.74 \\
& 22.1 & 151.1416225 & 41.2103033 & 2.74 \\
& 22.2 & 151.1471250 & 41.2152436 & 2.74 \\
& 3.1 & 151.1414121 & 41.2099250 & 3.288 \\
& 3.2 & 151.1476847 & 41.2152121 & 3.288 \\
& 31.1 & 151.1413250 & 41.2099825 & 3.288 \\
& 31.2 & 151.1477393 & 41.2151976 & 3.288 \\
& 32.1 & 151.1412104 & 41.2100544 & 3.288 \\
& 32.2 & 151.1478065 & 41.2151979 & 3.288 \\
& 33.1 & 151.1411279 & 41.2101547 & 3.288 \\
& 33.2 & 151.1478809 & 41.2151884 & 3.288 \\
& 33.3 & 151.1418864 & 41.2116948 & 3.288 \\
& 4.1 & 151.1439081 & 41.2165866 & 3.332 \\ 
& 4.2 & 151.1382517 & 41.2153846 & 3.332 \\  
& 4.3 & 151.1379048 & 41.2149959 & 3.332 \\  
& 4.4 & 151.1434099 & 41.2103752 & 3.332 \\  
& 41.1 & 151.1441118 & 41.2165193 & 3.332 \\  
& 41.2 & 151.1383309 & 41.2153297 & 3.332 \\  
& 41.3 & 151.1378932 & 41.2148820 & 3.332 \\  
& 41.4 & 151.1434562 & 41.2102573 & 3.332 \\  
& 42.1 & 151.1444522 & 41.2163884 & 3.332 \\  
& 42.2 & 151.1383940 & 41.2153469 & 3.332 \\  
& 42.3 & 151.1378407 & 41.2148091 & 3.332 \\  
& 42.4 & 151.1434818 & 41.2101761 & 3.332 \\  
& 43.1 & 151.1445319 & 41.2162919 & 3.332 \\  
& 43.2 & 151.1384506 & 41.2154232 & 3.332 \\  
& 43.3 & 151.1376594 & 41.2145747 & 3.332 \\  
& 43.4 & 151.1435603 & 41.2101349 & 3.332 \\   
& 43.5 & 151.1424833 & 41.2118271 & 3.332 \\ 

\SDSSthreeim & & & & \\
& QSO-A & 157.3081009 & 26.38830445 & 2.1992 \\ 
& QSO-B & 157.3093619 & 26.39446237 & 2.1992 \\  
& QSO-C & 157.3095755 & 26.3939894 & 2.1992 \\  
& 1.1 & 157.2980611 & 26.3907404 & \nodata \\
& 1.2 & 157.2978817 & 26.3924467 & \nodata \\
& 1.3 & 157.3008758 & 26.3974054 & \nodata \\
& 2.1 & 157.2981743 & 26.3915325 & 2.1812 \\
& 2.3 & 157.3014749 & 26.3977063 & 2.1812 \\
& 3.1 & 157.2990642 & 26.3923892 & 3.0275 \\
& 3.2 & 157.3074114 & 26.3913469 & 3.0275 \\
& 3.3 & 157.3041512 & 26.3982630 & 3.0275 \\
& 3.4 & 157.3015481 & 26.3880193 & 3.0275 \\  
& 3.5 & 157.3017377 & 26.3879213 & 3.0275 \\  
& 3.6 & 157.3018385 & 26.3878900 & 3.0275 \\  
& 3.7 & 157.3032208 & 26.3919632 & 3.0275 \\
& 4.1 & 157.2992278 & 26.3925219 & 3.0278 \\
& 4.2 & 157.3076382 & 26.3913247 & 3.0278 \\
& 4.3 & 157.3043869 & 26.3981437 & 3.0278 \\
& 4.4 & 157.3023985 & 26.3877048 & 3.0278 \\
& 4.5 & 157.3035100 & 26.3920169 & 3.0278 \\
& 5.1 & 157.3019777 & 26.3946563 & 1.0232 \\ 
& 5.3 & 157.3008781 & 26.3917377 & 1.0232 \\ 
& 7.1 & 157.3075794 & 26.3951262 & 5.0622 \\
& 7.2 & 157.3064130 & 26.3960500 & 5.0622 \\ 
& 7.3 & 157.3014210 & 26.3936610 & 5.0622 \\ 
& 7.4 & 157.3012420 & 26.3938020 & 5.0622 \\ 

\caption{Positions and spectroscopic redshifts of the multiply-imaged background sources used as constraints in the strong lens models for \SDSSfiveim\ and \SDSSthreeim.  See Table 1 from \cite{Sharon2017} for the lensing constraints for \SDSSsixim.}  
\label{table:constraints}
\end{longtable}
\end{center}

\begin{table}[h!]
\begin{center}
\begin{tabular}{c c c c c c c c}
\hline
Component No. & $\Delta$ R.A. [$''$] & $\Delta$ Decl. [$''$] & \textit{$e$} & \textit{$\theta$} [deg] & \textit{$\sigma_{0}$} [km s$^{-1}$] & \textit{r$_{cut}$} [kpc] & \textit{r$_{c}$} [kpc] \\
\hline
1 & 0.085$^{+0.52}_{-2.56}$ & 3.07$^{+5.83}_{-1.30}$ & 0.17$^{+0.022}_{-0.030}$ & 66.39$^{+3.70}_{-3.22}$ & 987$^{+245}_{-84}$ & [1500] & 126.27$^{+112.43}_{-33.97}$\\
2 & [0] & [0] & [0.40] & 63.98$^{+4.34}_{-5.31}$ & 461$^{+48}_{-52}$ & 181.42$^{+13.77}_{-28.04}$ & 5.65$^{+0.99}_{-1.62}$\\
3 & [-1.963] & [-1.832] & 0.42$^{+0.25}_{-0.19}$ & [349.480] & 235$^{+10}_{-14}$ & 30.30$^{+7.045}_{-12.29}$ & 2.68$^{+0.99}_{-0.68}$\\
4 & [-7.659] & [-9.821] & 0.43$^{+0.22}_{-0.29}$ & [131.13] & 127$^{+33}_{-29}$ & 20.13$^{+6.64}_{-8.33}$ & 1.62$^{+1.48}_{-1.06}$\\
5 & [8.463] & [-3.877] & 0.44$^{+0.24}_{-0.27}$ & [133.89] & 114$^{+31}_{-28}$ & 13.28$^{+2.97}_{-2.97}$ & 2.26$^{+0.92}_{-1.20}$\\
6 & [-11.220] & [11.401] & 0.42$^{+0.29}_{-0.29}$ & 150.24$^{+22.22}_{-34.44}$ & 76$^{+9}_{-7}$ & 22.46$^{+5.79}_{-6.85}$ & 3.18$^{+0.85}_{-0.85}$\\
\end{tabular}
\caption{Strong lensing mass model parameters for \SDSSfiveim.  Median values and the 1$\sigma$ confidence level from the MCMC are reported.  The coordinates $\Delta$ R.A. and $\Delta$ Decl. are listed in arcseconds measured east and north from the core of Component No. 2 at [RA, Dec] = [151.142381, 41.212131].  The other parameters are the ellipticity \textit{e}, the position angle \textit{$\theta$}, the velocity dispersion \textit{$\sigma_{0}$}, the cut radius \textit{r$_{cut}$}, and the core radius \textit{r$_{c}$}.  The position angle \textit{$\theta$} is measured north of west.  In converting from from angular units to kpc, an \Hnaut\ value of 70 \unitsHnaut\ was used.  The parameters listed in square brackets were not optimized.}
\label{table:1004_model}
\end{center}
\end{table}

\begin{table}[h!]
\begin{center}

\begin{tabular}{m{3cm} m{1cm} m{1cm} m{0.5cm} m{1cm} m{0.75cm} m{0.5cm} m{4cm} m{2.25cm}}
 \hline
 Target Name & \textit{z} cluster & \textit{z} QSO & no. QSO im & widest separation [$''$] & no. of lensed sources & no. of spec-zs &  time delay (days) & Reference \\ [0.5ex] 
 \hline
 \SDSSfiveim & 0.68 & 1.734 & 5 & 14.6 & 4 & 4 & $\Delta t_{AB} = -43.01 \pm 0.27$ & Mu$\tilde{n}$oz+(2022) \\ 
 & & & & & & & $\Delta t_{AC} = -825.23 \pm 0.46$ & \\ 
 & & & & & & & $\Delta t_{AD} = 1633.23 \pm 0.97$ & \\ 
 
 \SDSSthreeim & 0.58 & 2.1992 & 3 & 22.5 & 7 & 6 & $\Delta t_{AB} = 744 \pm 10$ & Fohlmeister+(2013) \\

 \SDSSsixim & 0.49 & 2.805 & 6 & 15.1 & 5 & 3 & $\Delta t_{AB} = 42.44^{+1.36}_{-1.44}$ & Dyrland (2019) \\
 & & & & & & & $\Delta t_{AC} = -696.65^{+2.00}_{-2.10}$ & \\
 \hline
\end{tabular}

\caption{The three large separation lensed QSOs in the \HST\ archive.  The listed time delays are the most up-to-date values from the literature.  See \cite{Fohlmeister2008} and \cite{Dahle2015} for previous measurements for \SDSSfiveim\ and \SDSSsixim, respectively.}
\label{table:time_delays}
\end{center}
\end{table}

\begin{table}[h!]
\begin{center}
\begin{tabular}{c c c c c c c c}
\hline
Component No. & $\Delta$ R.A. [$''$] & $\Delta$ Decl. [$''$] & \textit{$e$} & \textit{$\theta$} [deg] & \textit{$\sigma_{0}$} [km s$^{-1}$] & \textit{r$_{cut}$} [kpc] & \textit{r$_{c}$} [kpc]\\
\hline
1 & 10.01$^{+0.62}_{-0.53}$ & 0.71$^{+0.25}_{-0.23}$ & 0.53$^{+0.031}_{-0.034}$ & 172.80$^{+2.24}_{-2.27}$ & 650$^{+21}_{-20}$ & [1500] & 31.39$^{+4.37}_{-3.78}$\\
2 & -3.04$^{+1.38}_{-1.16}$ & 3.62$^{+0.46}_{-0.58}$ & 0.55$^{+0.052}_{-0.055}$ & 17.25$^{+4.87}_{-5.10}$ & 528$^{+30}_{-20}$ & [1500] & 37.95$^{+6.42}_{-6.62}$\\
3 & -2.48$^{+1.25}_{-1.35}$ & -0.11$^{+1.83}_{-2.35}$ & 0.61$^{+0.10}_{-0.062}$ & 45.57$^{+7.24}_{-9.24}$ & 385$^{+43}_{-52}$ & [1500] & 57.82$^{+9.47}_{-11.86}$\\
4 & [-3.808] & [-1.354] & 0.51$^{+0.19}_{-0.21}$ & 69.07$^{+19.26}_{-15.61}$ & 202$^{+20}_{-19}$ & 33.64$^{+7.88}_{-6.82}$ & 1.92$^{+0.52}_{-0.86}$\\
5 & [19.7] & [-8.8] & [0.0] & [0.0] & 169$^{+30}_{-24}$ & 89.94$^{+19.27}_{-19.47}$ & [0.0] \\
6 & 23.87$^{+0.11}_{-0.13}$ & 6.50$^{+0.14}_{-0.12}$ & 0.30$^{+0.29}_{-0.20}$ & 52.06$^{+26.58}_{-38.88}$ & 64$^{+7}_{-5}$ & 32.65$^{+11.13}_{-16.82}$ & 0.51$^{+0.30}_{-0.31}$\\
\end{tabular}
\caption{Strong lensing mass model parameters for \SDSSthreeim.  Median values and the 1$\sigma$ confidence level from the MCMC are reported.  The coordinates $\Delta$ R.A. and $\Delta$ Decl. are listed in arcseconds measured east and north from [RA, Dec] = [157.302047, 26.392209].  The other parameters are the ellipticity \textit{e}, the position angle \textit{$\theta$}, the velocity dispersion \textit{$\sigma_{0}$}, the cut radius \textit{r$_{cut}$}, and the core radius \textit{r$_{c}$}.   The position angle \textit{$\theta$} is measured north of west.  In converting from from angular units to kpc, an \Hnaut\ value of 70 \unitsHnaut\ was used.  The parameters listed in square brackets were not optimized.}
\label{table:1029_model}
\end{center}
\end{table}

\begin{table}[h!]
\begin{center}
\begin{tabular}{c c c c c c}
\hline
System & $\Delta$t$_{AB}$ & $\Delta$t$_{AC}$ & $\Delta$t$_{AD}$ & $\Delta$t$_{AE}$ & $\Delta$t$_{AF}$ \\
\hline
\SDSSfiveim & -11 & -783 & 1294 & 1776 & N/A \\
\SDSSthreeim & 1060 & 1054 & N/A & N/A & N/A \\
\SDSSsixim & 54 & -693 & 485 & 564 & 431 \\
\end{tabular}
\caption{Predicted time delay (in days) from the ‘best' lens model for each cluster.  The values are measured at the model-predicted locations of the quasar images, assuming \Hnaut$=70$ \unitsHnaut.}

\label{table:pred_time_delays}
\end{center}
\end{table}

\begin{table}[h!]
\begin{center}
\begin{tabular}{c c c}
\hline
System & \Hnaut\ (\unitsHnaut) & \Hnaut\ (\unitsHnaut)\\
& (from best model) & (mean $\pm$ 1$\sigma$)\\
\hline
\SDSSfiveim & &\\
AB & \HfiveimABbest & \HfiveimAB\\
AC & \HfiveimCAbest & \HfiveimCA\\
AD & \HfiveimADbest & \HfiveimAD \\
\SDSSthreeim & &\\
AB & \HthreeimBAbest & \HthreeimBA \\
\SDSSsixim & &\\
AB & \HsiximABbest & \HsiximAB \\
AC & \HsiximCAbest & \HsiximCA \\
\end{tabular}
\caption{\Hnaut\ constraints from the time delay measurements in \SDSSfiveim, \SDSSthreeim, and \SDSSsixim.  The middle column is the \Hnaut\ value from the ‘best' lens model for each cluster.  The right column lists the mean and 1$\sigma$ from the Gaussian distribution fit to the \Hnaut\ values determined from 500 random models drawn from the MCMC.}    
\label{table:Hnaut_constraints}
\end{center}
\end{table}

\begin{table}[h!]
\begin{center}
\begin{tabular}{c c c c}
\hline
Model & $\chi^{2}$ & dof & rms (arcsec) \\
\hline
\SDSSfiveim & 16.16 & 51 & 0.119 \\
\SDSSthreeim & 25.44 & 15 & 0.132 \\
\SDSSsixim & 8.18 & 1 & 0.08 \\
\end{tabular}
\caption{The minimum $\chi^{2}$, degrees of freedom (dof), and rms scatters in the source plane.} 
\label{table:goodness_of_fit}
\end{center}
\end{table}



\end{document}